\documentclass[letterpaper, 10pt]{article}
\usepackage[margin=2.5cm]{geometry}

\usepackage[usenames, dvipsnames]{color}
\usepackage[noadjust]{cite}
\usepackage{booktabs}
\usepackage{graphicx}
\usepackage{caption}
\usepackage{subcaption}
\usepackage{float}
\usepackage{amsmath}
\usepackage{algorithmicx}
\usepackage{subfiles}
\usepackage{multirow}
\usepackage{mathrsfs}
\usepackage{scalefnt}
\usepackage{listings}
\usepackage{tabularx}
\usepackage{longtable}
\usepackage{capt-of}
\usepackage{tikz}
\usepackage{pgfplots}
\usetikzlibrary{matrix}
\usepgfplotslibrary{groupplots}
\usepackage{float}
\usepackage[all,defaultlines=4]{nowidow}

\graphicspath{{images}{./media/}}

\usepackage{array}
\newcolumntype{L}[1]{>{\raggedright\let\newline\\\arraybackslash\hspace{0pt}}m{#1}}
\newcolumntype{C}[1]{>{\centering\let\newline\\\arraybackslash\hspace{0pt}}m{#1}}
\newcolumntype{R}[1]{>{\raggedleft\let\newline\\\arraybackslash\hspace{0pt}}m{#1}}

\usepackage{soul}
\usepackage[export]{adjustbox}

\usepackage[pdfencoding=auto, pdfusetitle,hidelinks]{hyperref}

\title{\LARGE \bf
The Role of Emotion in Problem Solving: First Results from Observing Chess
}

\author{Thomas Guntz$^{1}$, James L. Crowley$^{1}$, Dominique Vaufreydaz$^{1}$, Raffaella Balzarini$^{1}$,\\Philippe Dessus$^{1,2}$}

\date{
\small{$^{1}$Univ. Grenoble Alpes, CNRS, Inria, Grenoble INP, LIG, 38000 Grenoble, France\\
$^{2~}$Univ. Grenoble Alpes, LaRAC, 38000 Grenoble, France}\\
\vspace{0.5em}\textit{Author version}
}

\hypersetup{
  pdftitle={The Role of Emotion in Problem Solving: First Results from Observing Chess},
  pdfauthor={Thomas Guntz, James L. Crowley, Dominique Vaufreydaz, Raffaella Balzarini, Philippe Dessus},
  pdfkeywords={Working memory, Concept Formation, Chunking, Situation Modeling, Emotions, Problem Solving},
}

\begin{document}
\scalefont{0.995}
\maketitle
\thispagestyle{plain}
\pagestyle{plain}
\setlength{\belowdisplayskip}{2pt}
\widowpenalty10000
\clubpenalty10000
\addtolength{\abovedisplayskip}{-5pt}

\begin{abstract}

In this paper we present results from recent experiments that suggest
that chess players associate emotions to game situations and reactively
use these associations to guide search for planning and problem solving.
We describe the design of an instrument for capturing and interpreting 
multimodal signals of humans engaged in solving challenging problems.
We review results from a pilot experiment with human
experts engaged in solving challenging problems in Chess that revealed an unexpected
observation of rapid changes in emotion as players attempt to solve
challenging problems. We propose a cognitive model that describes the  
process by which subjects select chess chunks for use in interpretation of the game
situation and describe initial results from a second experiment designed to test this model.

\end{abstract}

\vspace{0.5em}
\begin{changemargin}{0.9cm}{0.9cm} 
\end{changemargin}
\vspace{0.5em}

\section{Introduction}

Humans display awareness and emotions through a variety of non-verbal
channels. It is increasingly possible to record and interpret such
information with available technology. Publicly available software can
be used to efficiently detect and track face orientation using web
cameras. Concentration can be inferred from changes in pupil size
\cite{kahneman1966pupil}. Observation of Facial Action Units \cite{ekman1971constants} can
be used to detect both sustained and instantaneous (micro-expressions)
displays of valence and excitation. Heart rate can be measured from the
Blood Volume Pulse as observed from facial skin color \cite{poh2011advancements}. Body
posture and gesture can be obtained from low-cost RGB sensors with depth
information (RGB+D) \cite{shotton2013real} or directly from images using
detectors learned using deep learning \cite{ramakrishna2014pose}. Awareness
and attention can be inferred from eye-gaze (scan path) and fixation
using eye-tracking glasses as well as remote eye tracking devices
\cite{holmqvist2011eye}. Such recordings can be used to reveal awareness of
the current situation and to predict ability to respond effectively to
opportunities and threats.

\begin{figure}
    \centering
    \includegraphics[width=2.61736in,height=2.61736in]{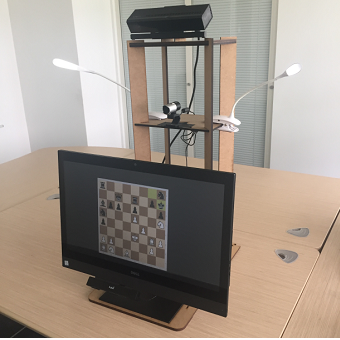}
    \caption{Instrument for recording gaze and emotions of experts engaged
in solving problems. On top, a Kinect2 oriented at the player. In the
middle, a USB camera to capture the face. Illumination is provided by
two LED lamps on the sides. A wooden structure is used to provide a
fixed position and orientation for all sensors.}
    \label{fig:setup}
\end{figure}

We have constructed an instrument for capturing and interpreting
multimodal signals of humans engaged in solving challenging problems.
Our instrument, shown in Figure \ref{fig:setup}, captures eye gaze, fixations, body
postures, and facial expressions signals from humans engaged in
interactive tasks on a touch screen. We use a 23 inch Touch-Screen
computer, a Kinect 2.0 mounted 35 cm above the screen to observe the
subject, a 1080p Webcam for a frontal view, a Tobii Eye-Tracking bar
(Pro X2-60 screen-based) and two adjustable USB-LED for lighting
condition control. A wooden structure is used to rigidly mount the
measuring equipment in order to assure identical sensor placement and
orientation for all recordings.

Several software systems have been used for recording and analyzing data
in our pilot experiment. Tobii Studio 3.4.7 was used for recording and
analysis of eye-gaze. Noldus FaceReader 7.0 was used for emotion
detection. Body posture was provided by both the Kinect 2.0 SDK and by
an enhanced version of a real-time multi-person pose estimation software
\cite{cao2017realtime}. This can be used to observe number of indicators for
stress from body posture including agitation, body volume and
self-touching. Body Agitation was computed from the intensity of joint
rotations. Body volume is the space occupied by the 3D bounding box
built around joints \cite{johal2015cognitive}. Self-Touching was determined from
collisions between wrist-elbow segments and the head, and is known to be
correlated with negative affect as well as frustration in problem
solving \cite{harrigan1985self}.

As a pilot study, we observed expert chess players engaged in solving
problems of increasing difficulty \cite{guntz2018a}. 
During the study, data were recorded from all sensors (Kinect 2, Webcam,
Screen capture, user clicks, Tobii-Bar) using the RGBD Sync SDK from the
Mobile RGBD project. This framework provides synchronization of all data
by associating a timestamp with a millisecond precision to each recorded
frame and can be used to read, analyze and display recorded data.
Our initial hypothesis
was that we could directly detect awareness of significant
configurations of chess pieces (chunks) from eye-scan and physiological
measurements of emotion in reaction to game situation. The pilot
experiment demonstrated that this initial hypothesis was naive. This
paper reports on our attempts to provide interpretations of those
results, as well as a more recent set of experiments aimed at confirming
our interpretation.

\section{Results from the First Pilot Experiment}

A first pilot experiment was designed to validate our instrument and to
evaluate the effectiveness of different systems and sensors for
observing eye-gaze, facial expressions, body posture, pupil dilation,
and cardiac rhythm. With the aid of the president of a local chess club,
we defined 11 end-game chess problems, similar to the daily chess
puzzles that can be found in magazines or on chess websites. Each of
these tasks challenges the subject to check-mate the opponent in a
number of predefined moves ranging from 1 to 6. Tasks requesting 1 to 3
moves are viewed as easy whereas tasks requiring 4 to 6 moves are
considered to be challenging. We ordered the problems in increasing
difficulty with 4 tasks with one move, 4 tasks with two and three moves
(2 of each) and 3 tasks with four, five and six moves to mate. The tasks
were presented with the subject alternately playing black or white. Full
results with eye-gaze and fixation are discussed in \cite{guntz2018a}. Here
we concentrate on observations of emotion.

\subsection{Data Collection}

Subjects were asked to solve chess tasks within a fixed, but
unspecified, time period. We recorded eye gaze, facial expressions, body
postures and physiological reactions of the players as they solved
problems of increasing difficulty. In a recording session with the local
chess club, we recorded 21 subjects solving the problems. Unfortunately
7 of the recordings were unusable because of problems with eye-tracking.
A few weeks later we had the opportunity to test 9 additional subjects
during a local tournament, including six casual players who were not
currently playing in club. These two sessions yielded a total of 22
useful recordings. All of our subjects have Elo ratings\footnote{Elo
  ratings are an open ended scale used to rate chess players based on
  tournament performance. The system is named after its creator, Arpad
  Elo, a Hungarian-American physics professor.}, with 9 experts (Elo
ratings from 1759 to 2150) and 14 intermediates level players (Elo 1100
to 1513).

Participants were tested individually in sessions lasting approximately
45 min. Subjects were made aware of the eye-tracking bar in order to
perform a calibration step. No other information was given about the
recording equipment. After providing informed consent from each subject,
the Lichess web platform was presented and participants were asked to
play a practice game against a weak opponent provided by the
Stockfish\footnote{Stockfish is an open-source game engine used in many
  chess software systems, including Lichess. Stockfish levels range from
  level 1 (easiest) to level 8 (most challenging).} game engine playing
at level 1. No recording was made during this first game.

Each subject was individually presented with the N-Check-Mate tasks. The
number of moves needed for the N-Check-Mate tasks was communicated to
the subject and subjects were informed of an unspecified time limit. The
time constraints range from 2 min for the easiest tasks (1--2 moves) to
5 min for the most challenging tasks (4--5--6 moves). Subjects knew only
that they were limited to a few minutes to solve the task, although an
announcement was made when only one minute was remaining. If the subject
was unable solve the task within the time limit, the task was considered
as failed and the experiment proceeded to the next task. The experiment
is considered complete once all tasks had been completed.

\subsection{Data Analysis}

The Noldus FaceReader software \cite{langner2010presentation} was used to obtain both
sustained and instantaneous (micro-expression) displays of emotion, as
well as cardiac rate using Blood Volume Pulse. FaceReader uses the Viola
Jones face detector \cite{viola2004robust} to locate the largest face in the
field of view in order to detect motion of 20 Facial Action Units. Each
action unit is assigned a score between 0 and 1 and these are used to
classify the subject's emotions into one of Ekman's six basic emotion
states\footnote{Happiness, sadness, anger, fear, disgust, surprise and
  neutral.}. We were able to use the basic emotions provided by Noldus to
estimate values for Valence and Arousal. Self-Touching is determined
from collisions between wrist-elbow segments and the head.

During the pilot experiment, we encountered many cases where the face
detector used by FaceReader failed, primarily due to head movements or
occlusions. Comparative experiments with OpenFace 2.0 \cite{baltrusaitis2018openface}
using data recorded in the pilot demonstrated that OpenFace
provided more reliable detection and tracking of faces as well as
reliable tracking of facial action units \cite{baltruvsaitis2015cross}. For
subsequent experiments we have moved to using Open Face 2.0.

\subsection{Results from the pilot experiment}

Figure 2 shows the number of self-touches and changes of emotion for
intermediate and expert players over our increasingly challenging
problem set. Our initial hypothesis was that subjects would exhibit
sustained displays of emotions ranging from pleasure to frustration as
the difficulty of the problems increased. We were surprised to observe
that this was not the case. Rather, both self-touching and rate of
change in emotion state evolved from a neutral emotion during reactive
play to a period of frequent touching and rapid changes in emotion as
the problems became more challenging.

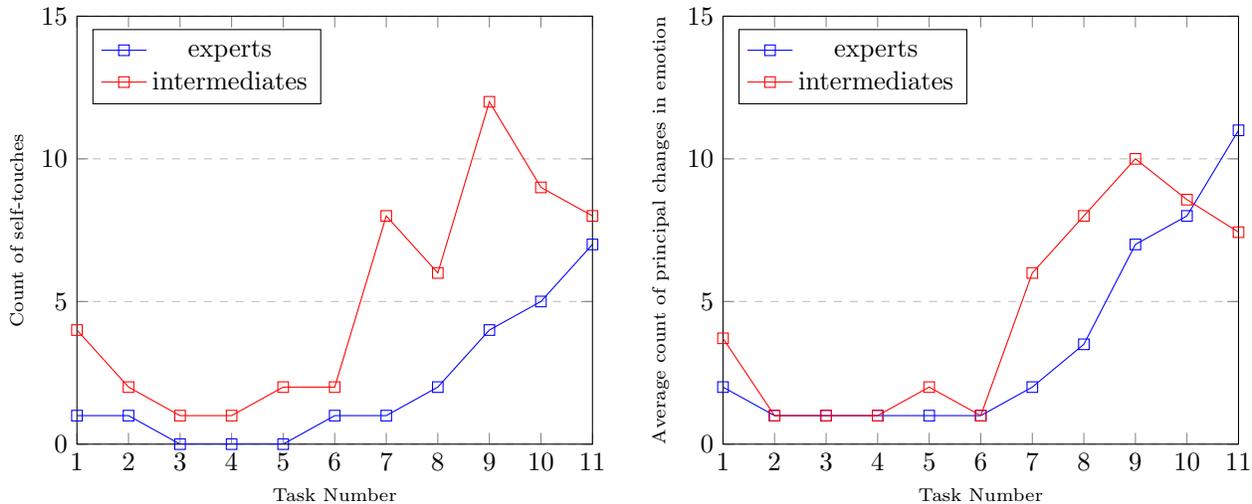
\begin{figure}
    \centering
\hspace{-2.5cm}
\begin{subfigure}[b]{0.35\linewidth}
\begin{tikzpicture}
\pgfplotsset{compat=1.14}{
   scale only axis,
    width=4cm,
    height=4cm, 
   compat=1.3
}
\begin{axis}[
    xlabel={},
    ylabel={},
    xmin=1, xmax=11, xtick={1,2,3,4,5,6,7,8,9,10,11},
    ymin=0, ymax=15,
    xlabel={\scriptsize{Task Number}},
    ylabel={\scriptsize{Count of self-touches}},
    legend pos=north west,
    ymajorgrids=true,
    grid style=dashed,
]
\addplot[
    color=blue,
    mark=square,
    ]
    coordinates {
    (1, 1)
    (2, 1)
    (3, 0)
    (4, 0)
    (5, 0)
    (6, 1)
    (7, 1)
    (8, 2)
    (9, 4)
    (10, 5)
    (11, 7)
    };
    
\addplot[
    color=red,
    mark=square,
    ]
    coordinates {
    (1, 4)
    (2, 2)
    (3, 1)
    (4, 1)
    (5, 2)
    (6, 2)
    (7, 8)
    (8, 6)
    (9, 12)
    (10, 9)
    (11, 8)
    };    
\legend{experts, intermediates}
\end{axis}
\end{tikzpicture}
\label{fig:results:previous_analysis}
\end{subfigure}
\hspace{2.5cm}
\begin{subfigure}[b]{0.35\linewidth}
\begin{tikzpicture}
\pgfplotsset{compat=1.14}{
   scale only axis,
    width=4cm,
    height=4cm, 
   compat=1.3
}
\begin{axis}[
    xlabel={},
    ylabel={},
    xmin=1, xmax=11, xtick={1,2,3,4,5,6,7,8,9,10,11},
    ymin=0, ymax=15,
    xlabel={\scriptsize{Task Number}},
    ylabel={\scriptsize{Average count of principal changes in emotion}},
    legend pos=north west,
    ymajorgrids=true,
    grid style=dashed,
]

    \addplot[
    color=blue,
    mark=square,
    ]
    coordinates {
    (1, 2)
    (2, 1)
    (3, 1)
    (4, 1)
    (5, 1)
    (6, 1)
    (7, 2)
    (8, 3.5)
    (9, 7)
    (10, 8)
    (11, 11)
    };
    
\addplot[
    color=red,
    mark=square,
    ]
    coordinates {
    (1, 3.71)
    (2, 1)
    (3, 1)
    (4, 1)
    (5, 2)
    (6, 1)
    (7, 6)
    (8, 8)
    (9, 10)
    (10, 8.57)
    (11, 7.43)
    };    
\legend{experts, intermediates}
\end{axis}
\end{tikzpicture}
\label{fig:results:analysis}
\end{subfigure}
\vspace{-0.5cm}
    \caption{Self-touches (left) and average count of number of changes in
emotion state (right) for intermediate and experts over the 11 tasks.}
    \label{fig:my_label}
\vspace{-0.25cm}
\end{figure}

Figure 2 illustrates that the rate of changes of emotional state
increases with the difficulty for both intermediates and experts, with
significantly higher numbers for intermediate players. The correlation
with the rise in self-touching, confirms that subjects were increasingly
challenged. We conclude that frustration for intermediate players rose
rapidly for tasks 7, 8, 9 and 10, and then dropped, as subjects seemed
to abandon efforts to solve task 11. For experts, self-touching and
changes in emotion gradually increased for problems 7 through 11,
indicating that experts experienced only minor discomfort for these
problems.

\section{A Cognitive Model for Reasoning about Chess}

Our initial hypothesis was that rapid changes of emotion correspond to
success or failure of alternative branches during game tree exploration.
We now believe that this explanation is overly simplistic. Even expert
players are unable to hold the entire game state in working memory
\cite{Gobet1996}. The selection of the partial game description to hold in
working memory is critical for reasoning about chess.

In order to better understand the phenomena observed in our pilot
experiment, we have constructed a model of the cognitive processes
involved, using theories from cognitive science and classic (symbolic)
artificial intelligence. This model is a very partial description that
allows us to ask questions and make predictions to guide future
experiments. Our model posits that experts reason with a situation model
that is strongly constrained by limits to the number of entities and
relations that may be considered at a time. This limitation forces
subjects to construct abstract concepts (chunks) to describe game play,
in order to explore alternative moves. Expert players retain
associations of situations with emotions in long-term memory. The rapid
changes in emotion correspond to recognition of previously encountered
situations during exploration of the game tree. Recalled emotions guide
selection of situation models for reasoning. This hypothesis is in
accordance with Damasio's Somatic Marker hypothesis, which posits that
emotions guide behavior, particularly when cognitive processes are
overloaded \cite{damasio1991somatic}.

\subsection{Situation Models}

Situation models \cite{Johnson-Laird:1989:MM:102953.102965} provide a formal framework for
describing human comprehension and problem solving. In logic terms, a
situation model is a state graph, in which each state (situation) is
defined as a logical expression of predicates (relations) defined over
entities. Entities can represent observed phenomena as well as instances
of concepts, procedures or episodes from long-term memory. A change in
the relation between entities results in a change in situation.
Procedures and actions may be associated with situations for use in
planning and reasoning \cite{barraquand2008learning}. Consequences may also be
associated with situations, making it possible to anticipate
opportunities and threats.

Relations describe associations of entities. Examples include spatial or
temporal order relations \cite{allen1983maintaining}, as well as ownership and family
or social associations. The number of arguments of a relation is called
its arity. Formally, a relation can be defined with any number of
arguments, including zero. In semantic web tools such as RDF and OWL,
arity 2 predicates are used to relate a subject to a property (an
object), expressed with the following schema: (Subject Relation Object).
However, in many domains, including chess, it is necessary to use
relations with different arities.

\subsection{Concept Schema for Entities and Relations}

Formal representations of concepts can be expressed as frames \cite{Minsky1974}. Frames define abstract concepts that can be instantiated as
entities. A frame associates the entity class with a set of properties,
and a set of procedures. Formally, the properties and procedures
associated with a concept can be seen as a form of relation \cite{sowa2000knowledge}. However, the association of properties within a concept is
internal and immutable. The values may change, but the set of properties
and procedures are fixed. Changing the set of properties or procedures
generates a new concept.

The basic entities in chess are the individual pieces with properties
that can include the type of piece, color (white or black), and board
position. This can be defined as a frame using the following schema:

\begin{minipage}{\textwidth}
\begin{center}
\small{\begin{lstlisting}[keepspaces=true]
(ChessPiece (piece-ID)
    (kind (one-of (king, queen, bishop, knight, rook, pawn)))
    (color (one-of (black white)))
    (position (row (range 1 to 8) (column (range a to h))
    (actions (move-procedure))
)
\end{lstlisting}}
\end{center}
\end{minipage}

Recognition of a concept creates an instance of the concept as an entity
in the situation model. Entities are labeled with a unique identity
(piece-ID) that can be used as a reference in instantiating relations.
Move-procedure is a procedure for generating the set of legal moves for
the piece.

The board configuration can be understood as a collection of relations
in which pieces threaten pieces of the opposing color and defend pieces
of the same color. As with entities, relations are instances of abstract
concepts defined as frames. A concept schema for a binary (arity-2)
chess relation would be:

\begin{minipage}{\textwidth}
\begin{center}
\small{\begin{lstlisting}[keepspaces=true]
(Relation (Relation-ID)
    (Name (relation-name))
    (Kind (one-of (offensive, defensive)))
    (Subject (entity-ID))
    (Object (entity-ID))
)
\end{lstlisting}}
\end{center}
\end{minipage}

Relations can be defensive, in which a piece protects or defends another
piece of the same color, or offensive in which a piece threatens a piece
of an opposing color. The subject and object are pointers to instances
of pieces or chunks that are held as entities in situation model.

A concept schema for an arity-3 relation would be:

\begin{minipage}{\textwidth}
\begin{center}
\small{\begin{lstlisting}[keepspaces=true]
(Relation (Relation-ID)
    (Name (relation-name))
    (Kind (one-of (offensive, defensive)))
    (Subject (entity-ID))
    (Object1 (entity-ID))
    (Object2 (entity-ID))
)
\end{lstlisting}}
\end{center}
\end{minipage}

\begin{center}
\begin{longtable}[]{@{}ccc@{}}
\toprule
\includegraphics[height=0.225\linewidth]{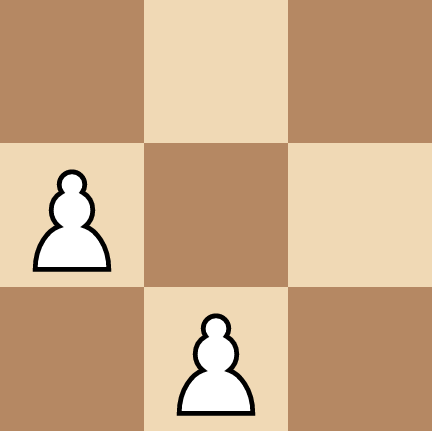}
& \includegraphics[height=0.225\linewidth]{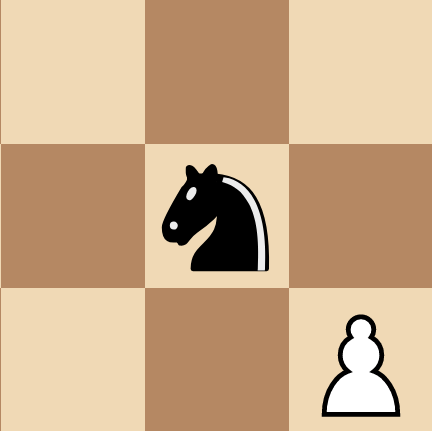}
&
\includegraphics[height=0.225\linewidth]{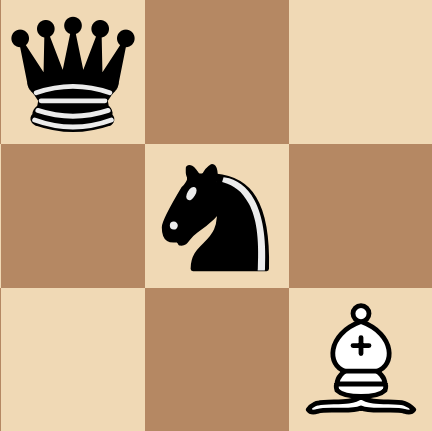}\tabularnewline
\midrule
a) protects &	b) threatens	& c) pins
\endhead
\bottomrule
\end{longtable}
\vspace{-0.5cm}
  \captionof{figure}{Three examples of relations. a) A defensive binary relation:
(pawn protects pawn) b) An offensive binary relation: (bishop threatens
knight) c) an offensive ternary relation: (bishop pins knight to queen)}%
\label{fig:chess_relation}
  \addtocounter{table}{-1}%
\end{center}

Figure \ref{fig:chess_relation} shows examples of chess relations. Figure 3a is a binary
defensive relation "protects", while 3b is a binary offensive relation
"threatens". Both of these can be expressed in terms of the classic
binary relation schema (subject relation object). Figure 3c shows a
ternary (arity-3) relation: "pins" with a subject and two objects.

Note that for every binary relation there is an inverse relation where
the object and subject roles are switched. For example (pawns threatens
knight) implies an inverse relation (knight threatened-by pawn). For
arity-3 relations, there are 6 inverse relations, one each for each
possible permutation of the subject, object1, and object2. Each relation
constrains possible moves by the subject piece.

\subsection{Working Memory}

Limits on the size of working memory are a fundamental property of human
cognition. In a seminal paper, G.A. Miller demonstrated that humans
could simultaneously retain between 5 and 9 things in short-term working
memory \cite{miller1956magical}. Most authors present working memory (WM) as a
collection of buffers that represent propositions representing perceived
phenomena or associated concepts from long-term memory (LTM). Cowan
\cite{cowan2014working}, reviews the history of research on working memory,
discussing the variety of definitions and experimental demonstrations
that have been used to describe this phenomena. Many authors report that
human adults are limited to 3 or 4 meaningful modal percepts (visual,
auditory, tactile, or propositional) and that unrehearsed percepts decay
within about 30 seconds.

Working memory is generally modeled as a form of Hebbian active memory
\cite{hebb1949organization}, \cite{linhares2005active} with propagation of activation energy
providing associations with other working memory elements, perception,
actions, and long-term memory. As activation energy spreads from WM into
LTM, multiple paths can lead to accumulations within cognitive units in
LTM \cite{anderson1983spreading}. When activation energy for a cognitive unit in
LTM exceeds the activation of an entity in WM, the LTM unit replaces the
contents of the WM unit. This can potentially impose a third limit on
working memory: a limit on the number of associations for an entity, as
excessive associations can dissipate the activation energy of a WM unit.

We note that activation energy may not be the only mechanism involved in
determining working memory contents. Cognitive skills such as mental
arithmetic or chess may involve automatically scanning through a set of
possible situation models in a predefined sequence.

\subsection{Representing Chess Concepts with Chunks}

The size of working memory limits the complexity of the description of
game state that a subject can retain and recall. Beginning players tend
to see the game in terms of the situations of individual pieces. Viewed
this way, the game state rapidly becomes extremely complex, as a large
number of pieces generate an even larger number of defensive and
offensive relations. Beginning players are blinded to possible
opportunities and threats by an overwhelming flood of information.
Learning to play involves learning to see the game in terms of
configurations of pieces, referred to as chunks.

Chunks are long-term memory constructs used for perception, recall and
reasoning about chess configurations. Figure 4 gives three examples of
chess chunks. The term was originally proposed in 1946 by De Groot \cite{de1978thought} as a form of cognitive unit used for reasoning by chess
masters. Each chunk is a distinct concept that can be directly
recognized by a player and represented as a unique entity in working
memory. Chunks allow players to accommodate the limits imposed by
working memory.

Chase and Simon \cite{chase1973perception} explored the use of chunks for perception
and recall by players of different levels, and demonstrated the
existence of chunks with memory experiments in which players with
different levels of expertise were required to remember configurations
of pieces. Simon and Gilmartin \cite{simon1973simulation} estimate that expertise in
chess may require between 10,000 and 100,000 chunks in long-term memory.
This rich vocabulary enables players to determine good moves with only
moderate search of the game tree.

\begin{center}
\begin{longtable}[]{@{}ccc@{}}
\toprule
\includegraphics[width=0.3\linewidth]{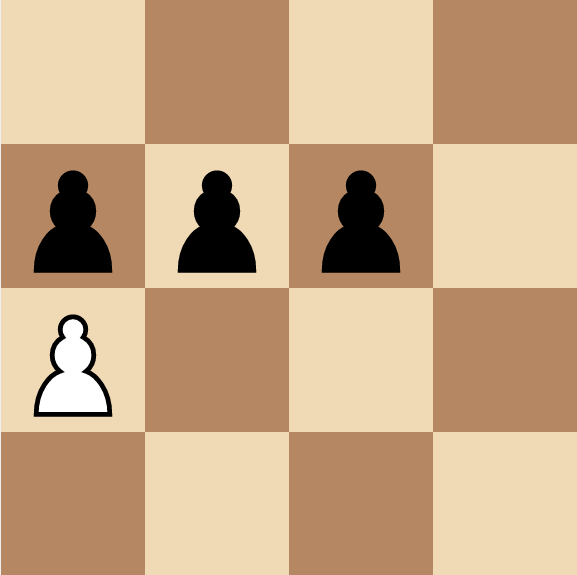}
& \includegraphics[width=0.3\linewidth]{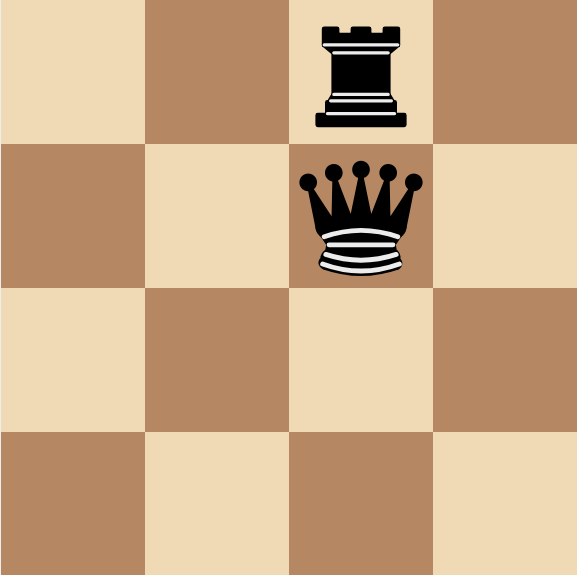}
& \includegraphics[width=0.3\linewidth]{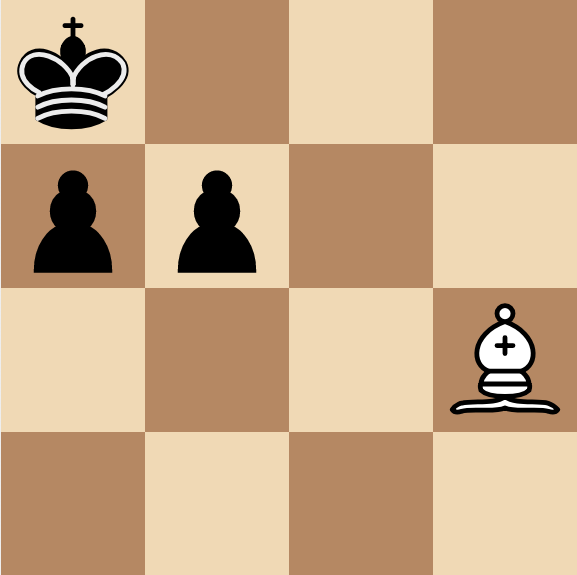}\tabularnewline
\midrule
a) wall of pawns & b) battery & c) trapped king
\endhead
\bottomrule
\end{longtable}
\vspace{-0.5cm}
  \captionof{figure}{Three examples of chess chunks. a) a wall of black
pawns. b) a battery in which a rook defends an
attacking queen. c) A trap in which a bishop prevents motion by the opposing king
offering a potential checkmate if a second threatening piece and be
added to the situation.}%
\label{fig:chess_chunk}
  \addtocounter{table}{-1}%
\end{center}

\subsection{A Situation Model for Chess}

The following is a possible schema for chess situations. Entities may be
instances of pieces or chunks. The actions slot provides a list of
possible moves that are enabled or prevented by the situation. The
emotion slot provides values for valence, arousal and dominance acquired
from experience and used to guide reasoning.

\begin{minipage}{\textwidth}
\begin{center}
\small{\begin{lstlisting}[keepspaces=true]
(ChessSituation (Situation-ID)
    (color (one-of (black, white)))
    (relations (Relation-ID*))
    (entities (Any-of (Piece-ID, Chunk-ID)*)
    (moves (moves*))
    (emotion (valence, arousal, dominance))
)
\end{lstlisting}}
\end{center}
\end{minipage}

The symbol * indicates a list of one of more elements. The number of
entities and relations within a situation is limited by the number or
elements in working memory. This number should not exceed Miller's limit
and is likely to be much smaller as part of working memory will be
consumed by other information about the game context such as imminent
threats or time constraints. For example, when considering consequences
of a move, the player must retain the current situation, the candidate
move and the resulting situation in working memory, as well as other
information about the game context. This would seem to limit the game
situation model to 3 or at most 4 entities. For example, a typical
situation model may be composed of one defending chunk, one attacking
chunk and a single active piece.

De Groot proposed that chess reasoning consists of 4 stages:
Orientation, Exploration, Investigation and Validation. Scan paths
observed in our experiments tend to confirm these phases. During the
orientation phase, players perceive chunks that can be used to model the
situation. Beginners tend to fixate on individual pieces, while expert
players can be observed to fixate on the center of chunks, but make only
very rapid saccades in the direction of component pieces. Small subsets
of chunks are then selected and used to explore possible moves. This
raises the question of how the player can select from among the large
available set of chunks. We believe that emotion plays a critical role
in such selection.

\section{The Role of Emotion in Reasoning}

Our current hypothesis is that the rapid changes in emotion displayed by
subjects in our pilot experiment are an involuntary display in reaction
to recognition of previously encountered situations during exploration
of the game state. With this interpretation, the player rapidly
considers partial descriptions as situations composed of a limited
number of perceived chunks. Recognition of situations from experience
evokes emotions that are displayed as face expressions and body posture.
Our hypothesis is that the subject uses the evoked emotions to select
from the many possible situations for reasoning about moves during
orientation and exploration.

\subsection{Observing and Modeling Human Emotions}

Ekman {[}Ekman 77{]} observed that emotions are expressed by coordinated
temporal activations of 21 different facial muscles assisted by a number
of additional muscles. Activations of these muscles are visible through
subtle deformations in the surface structure of the face. Ekman proposed
a formal coding system (Facial Action Coding or FACS) that can be
learned by humans and has been used to produce relatively reliable
software to detect facial muscle activations.

Ekman also proposed that six emotion states (joy, sadness, anger,
disgust, fear and surprise) are universal across cultures {[}Ekman
92{]}. As an alternative to recognizing emotions as states, a number of
investigators have proposed that emotions can be observed as
physiological parameters \cite{russell1977evidence}. The circumplex model \cite{russell1980circumplex} represents emotions as trajectories in a circular space with two
dimensions: arousal and valence. The center of the circle represents a
neutral valence and a medium level of arousal. A third dimension is
required to discriminate between anger and fear. Russel and Mehrabian
refer to this as dominance. This is closely related to frustration as
measured by self-touching as measured in our experiments.

\subsection{Emotions as a Guide for Reasoning in Chess}

Valence, arousal and dominance can serve to guide reasoning in chess.
These three values can be learned from experience and associated with
chess situations in long-term memory.

Dominance corresponds to the degree of experience with the recognized
situation. As players gain experience with alternate outcomes for a
situation, they become more assured in their ability to spot
opportunities and avoid dangers. Valence corresponds to whether the
situation is recognized as favorable (providing opportunities) or
unfavorable (creating threats). Arousal corresponds to the imminence of
a threat or opportunity.

With this model, a defensive player will give priority to reasoning
about unfavorable situations and associated dangers. An aggressive
player will seek out high valence situations. All players will give
priority to situations that evoke strong arousal. The amount of effort
that player will expend exploring a situation can determined by
dominance.

\section{A Second Experiment to Develop our Model}

In order to develop our model, we have conducted a second experiment in
which we players were asked to explain their reasoning. The objectives
were to determine if eye-gaze, valence, arousal and frustration could be
correlated with the four phases of reasoning proposed by De Groot, and
to construct an ontology for chess concepts (chunks and relations) used
by players.

These experiments were performed using OpenFace 2.0 for recording Facial
Action Units (AUs) and the EyeWorks software for measuring pupil
dilation. Ekman's basics emotions were computed from the AUs following
Ekman's formula \cite{ekman1971constants}. Valence was computed as the intensity of
AUs involved in positive emotions, minus intensity of AUs involved in
negative emotions. Arousal was computed as a combination of AUs averaged
over the last 60 seconds.

We considered two techniques for self-reporting: Concurrent Verbal
Analysis (CVA) and Retroactive Task Explanation (RTE) \cite{Kuusela00}.
With CVA, subjects are asked to think aloud and to explain their
reasoning as they solve the task. With RTE, subjects are asked to explain
their reasoning after completing each task. Both
techniques have possible problems. With CVA a part of working memory is
devoted to constructing the explanation, reducing the cognitive
resources available for resolving the task. With RTE, the subject may
invent an explanation that does not accurately reveal the cognitive
process that was used. After a short pilot experiment with 4 subjects,
we elected to use RTE.

For our second experiment we recruited 23 subjects (2 experts, 19
intermediates and 2 beginners). Expert players were active players with
Elo ratings from 1930 to 2000. For the intermediate players, the Elo
ratings ranged from 1197 to 1700. Twelve of the intermediates were
casual players who were not currently playing in club.

Subjects were initially asked to play two easy practice games to become
familiar with the equipment. We then recorded eye-gaze, emotional state,
pupil size, valence, arousal, and self-touching as subjects solved a
series of 7 tasks composed of 4 Mate-in-N tasks and 3 survival tasks.
The Mate-in-N tasks involved a diverse set of concepts, with some game
situations unbalanced in favor of the opponent, forcing conservative
players to focus on defensive moves. The 3 survival tasks were all
hopeless positions, where experts would generally prefer to resign. On
completion of each task, subjects were asked to explain their
understanding of the board situation, and the reason for their moves. We
specifically asked them to identify opportunities, threats and possible
moves that were considered, including those not taken.

Although we have only recently begun analysis of this data, some
interesting phenomena are evident. The following are examples from two
expert players who provided particularly clear explanations.

\newcommand{\GroupImageWidth}{0.3\linewidth}

\begin{figure}[]
\begin{longtable}[]{@{}ccl@{}}
\cmidrule[0.75pt]{1-2} 
\includegraphics[width=\GroupImageWidth]{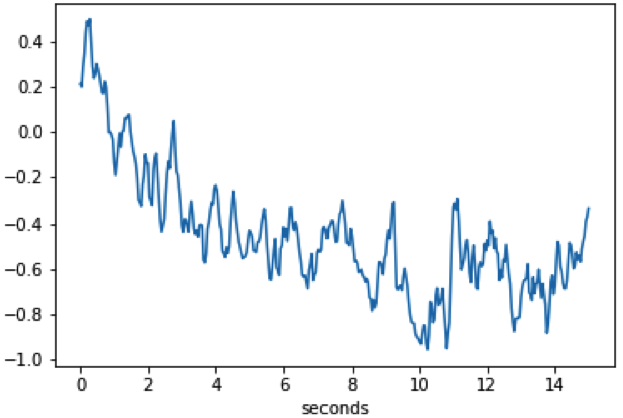}
&
\includegraphics[width=\GroupImageWidth]{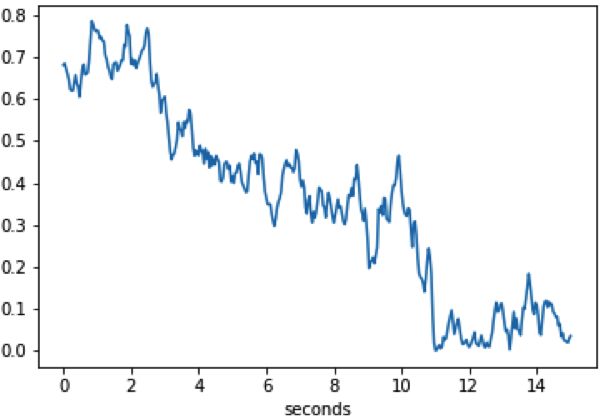}\tabularnewline
\cmidrule[0.5pt]{1-2} 
a) Valence & b) Arousal\tabularnewline
\cmidrule[0.75pt]{1-3} 
\includegraphics[width=\GroupImageWidth]{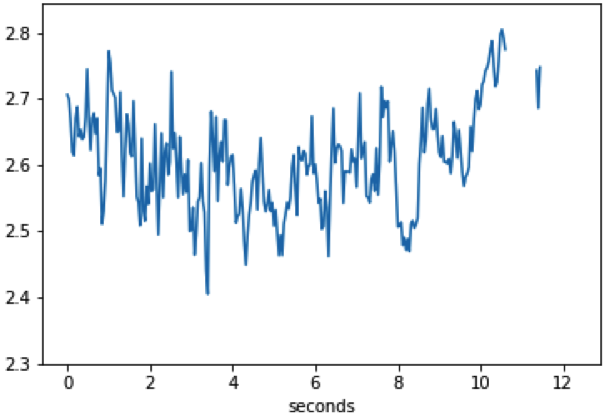}
&
\includegraphics[width=\GroupImageWidth]{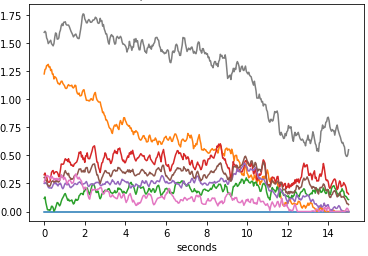} & \hspace{-0.4cm}\includegraphics[scale=0.43]{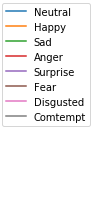}\tabularnewline
\cmidrule[0.5pt]{1-2} 
c) Pupil Diameter & d) Emotion States\tabularnewline
\cmidrule[0.75pt]{1-3}
\end{longtable}
\caption{Subject Q6 (expert ELO 1950+) solving task 4 (medium
difficulty). The subject rapidly recognized and confirmed the solution,
a shown by a steady decrease in valence, arousal and contempt. A rise in
pupil size is evident during the final, confirmation phase.}
\label{fig:subject_q6_task4}
\end{figure}

Figure \ref{fig:subject_q6_task4} on page~\pageref{fig:subject_q6_task4} shows an expert (subject Q6, ELO 1950+) resolving task 4
(moderately difficult mate-in-N) The subject rapidly recognized and
confirmed the solution as is evident in a steady decrease in valence,
arousal and contempt. A temporary rise in pupil size is evident during
the final, confirmation phase.

\begin{figure}[]
\begin{longtable}[]{@{}ccl@{}}
\cmidrule[0.75pt]{1-2} 
\includegraphics[width=\GroupImageWidth]{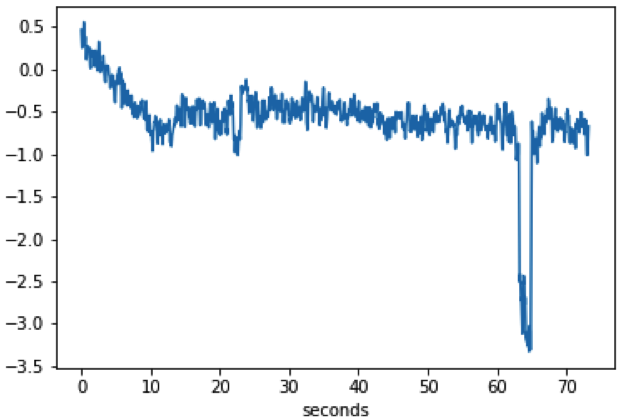}
&
\includegraphics[width=\GroupImageWidth]{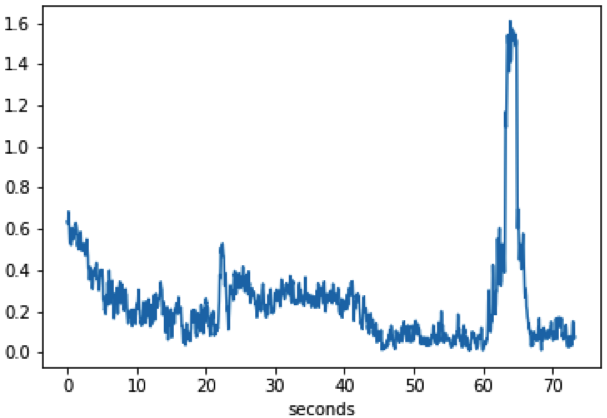}\tabularnewline
\cmidrule[0.5pt]{1-2} 
a) Valence & b) Arousal\tabularnewline
\cmidrule[0.75pt]{1-3} 
\includegraphics[width=\GroupImageWidth]{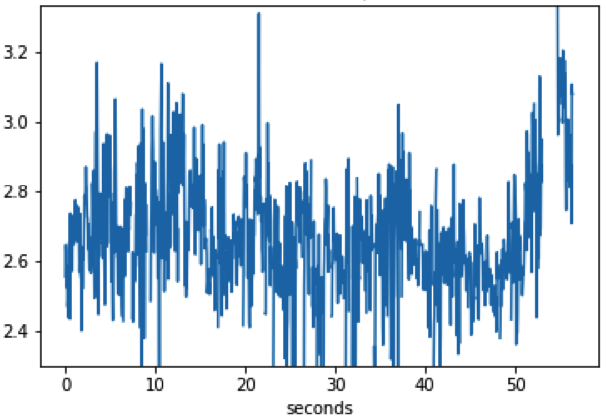}
&
\includegraphics[width=\GroupImageWidth]{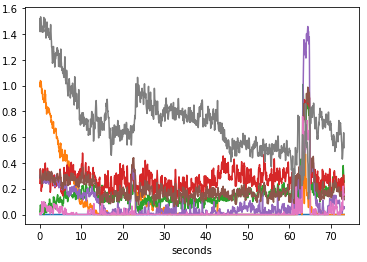}& \hspace{-0.4cm}\includegraphics[scale=0.43]{./media/legende2.png}\tabularnewline
\cmidrule[0.5pt]{1-2} 
c) Pupil Diameter & d) Emotion States\tabularnewline
\cmidrule[0.75pt]{1-3}
\end{longtable}
\caption{Subject Q6 (expert ELO 1950+) attempting to resolve a
Task 8 (Very difficult). The spike in arousal, valence, pupil diameter
and emotion states of fear and disgust correspond to a self-reported
recognition that the situation was hopeless.}
\label{fig:subject_q6_task8}
\end{figure}

Figure \ref{fig:subject_q6_task8} (page~\pageref{fig:subject_q6_task8}) shows the same expert (Q6, ELO 1950+) addressing the extremely
difficult task 8. The subject displays a sustained period of moderate
valence and low arousal and decreasing contempt, followed by a steep
drop in valence, a rapid spike in arousal and pupil size, and a sudden
peak in disgust and fear as the subject recognizes that there is no good
solution to the problem.

\begin{figure}[]
\begin{longtable}[]{@{}ccl@{}}
\cmidrule[0.75pt]{1-2} 
\includegraphics[width=\GroupImageWidth]{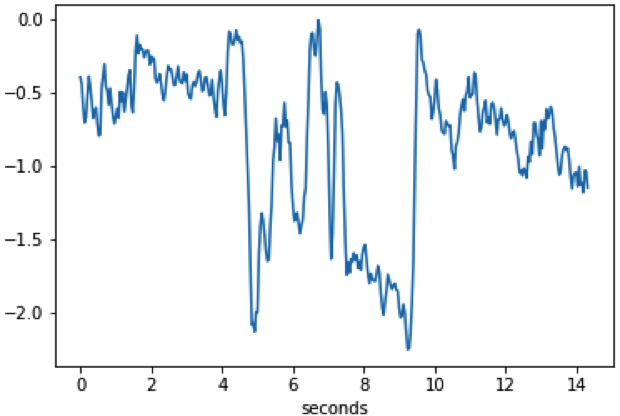} &
\includegraphics[width=\GroupImageWidth]{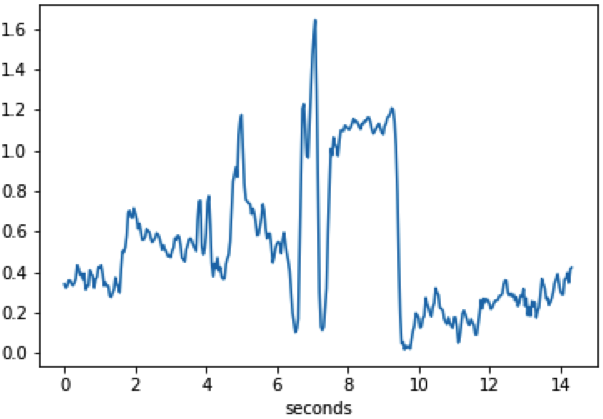}\tabularnewline
\cmidrule[0.5pt]{1-2} 
a) Valence & b) Arousal\tabularnewline
\cmidrule[0.75pt]{1-3} 
\includegraphics[width=\GroupImageWidth]{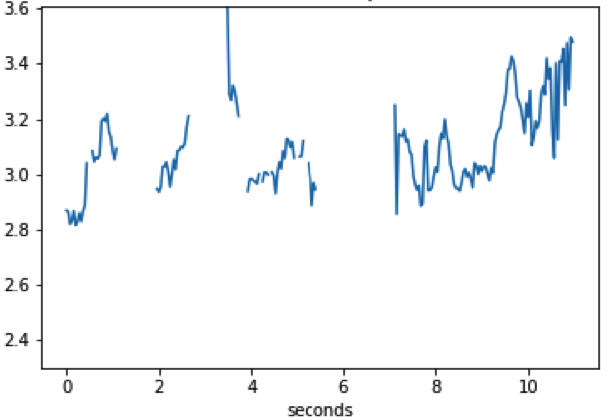}
&
\includegraphics[width=\GroupImageWidth]{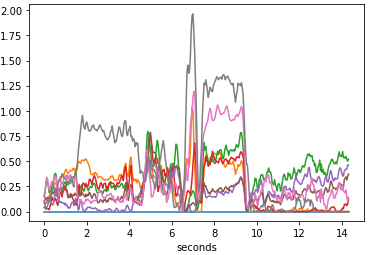} & \hspace{-0.4cm}\includegraphics[scale=0.43]{./media/legende2.png}\tabularnewline
\cmidrule[0.5pt]{1-2} 
c) Pupil Diameter & d) Emotion States\tabularnewline
\cmidrule[0.75pt]{1-3}
\end{longtable}
\caption{Subject Q12 (expert ELO 2000) attempting to resolve
Task 4. The moment where the subject discovers the solution is visible
as a strong correlation in valence, arousal, and contempt as well as a
self-report of the solution.}
\label{fig:subject_q12_task4}
\end{figure}

Figure \ref{fig:subject_q12_task4} (page~\pageref{fig:subject_q12_task4}) shows a different expert (Q12, ELO 2000+) solving the
moderately challenging Task 4. The moment where the subject recognizes
the solution coincides with a strong correlation in valence, arousal,
and contempt as well as a self-report of the solution. This is followed
by a second, less intense period of increasing valence, arousal and
contempt as the subject confirms the solution.

\begin{figure}[]
\begin{longtable}[]{@{}ccl@{}}
\cmidrule[0.75pt]{1-2} 
\includegraphics[width=\GroupImageWidth]{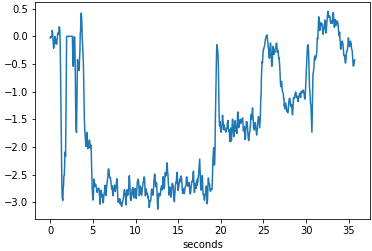}
&
\includegraphics[width=\GroupImageWidth]{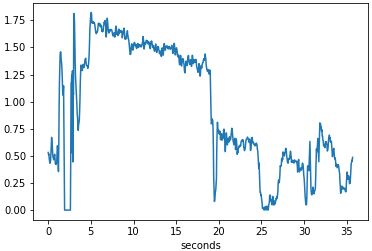}\tabularnewline
\cmidrule[0.5pt]{1-2} 
a) Valence & b) Arousal\tabularnewline
\cmidrule[0.75pt]{1-3}
\includegraphics[width=\GroupImageWidth]{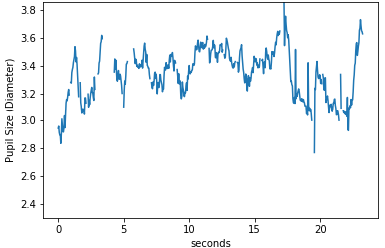}
&
\includegraphics[width=\GroupImageWidth]{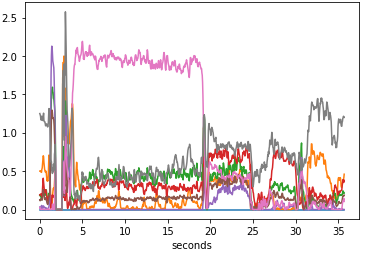} & \hspace{-0.4cm}\includegraphics[scale=0.43]{./media/legende2.png}\tabularnewline
\cmidrule[0.5pt]{1-2}
c) Pupil Diameter & d) Emotion States\tabularnewline
\cmidrule[0.75pt]{1-3}
\end{longtable}
\caption{Expert Q12 (Elo 2000+) confronted with the hopeless
situation of Task 8. The subject is visibly unhappy (strong negative
valence), very excited (strong arousal), and disgusted until recognizing
that the situation is hopeless after around 20 seconds.}
\label{fig:subject_q12_task8}
\end{figure}

Figure \ref{fig:subject_q12_task8} (page~\pageref{fig:subject_q12_task8}) shows subject Q12 (Elo 2000+) confronted with the hopeless
situation of Task 8. The subject is visibly unhappy (strong negative
valence), very excited (strong arousal), and disgusted until recognizing
that the situation is hopeless after around 20 seconds. The rise in
valence and drop in arousal can be interpreted as satisfaction as having
successfully understood the game situation.

\section{Conclusions}

Results from our initial experiment with recording eye-gaze and emotion
of chess experts showed an unexpected rapid variation of emotional state
as experts solved challenging problems. In this paper we have proposed a
model that explains this phenomena as an involuntary display of emotions
associated with recognition of situations. Our model suggests that an
association of emotions with recognized situations guides experts in
their selection of partial game configurations for use in exploring the
game tree. However, this is very much a work in progress, based on only
limited data.

We have presented initial results from a follow-on experiment designed
to explore the fidelity of our model, and to search for evidence of the
role of emotion in solving challenging problems. Initial results from
this second experiment appear to confirm our model. Further analysis and
additional experiments are needed to more confidently model the role of
emotions in reasoning.

\pagebreak
\section*{ACKNOWLEDGMENT}

This research has been be funded by the French ANR project CEEGE
(ANR-15-CE23-0005), and was made possible by the use of equipment
provided by ANR Equipment for Excellence Amiqual4Home
(ANR-11-EQPX-0002). Access to the facility of the MSH-Alpes SCREEN
platform for conducting the research is gratefully acknowledged. We are
grateful to all of the volunteers who generously gave their time to
participate in this study and to Lichess webmasters for their help and
approval to use their platform for this scientific experience. We would
like to thank Isabelle Billard, current chairman of the chess club of
Grenoble ''L'\'Echiquier Grenoblois`` and all members who participated
actively in our experiments. We also thank Professor Thomas Schack and
Thomas Küchelmann of Univ. Bielefeld Dept of Neuro-Cognition, for
valuable discussions and aid.

\bibliographystyle{plain}
\bibliography{references}
\end{document}